# Comparison of window shapes and lengths in short-time feature extraction for classification of heart sound signals

**Mahmoud Fakhry, Abeer FathAllah Brery**
Electrical Engineering Department, Faculty of Engineering, Aswan University, Aswan, Egypt

| Article Info | ABSTRACT |
|---|---|
| *Article history:*<br><br>Received Sep 16, 2021<br>Revised Jun 27, 2022<br>Accepted Jul 23, 2022<br><br>*Keywords:*<br><br>Bidirectional long short-term memory network<br>Classification<br>Feature extraction<br>Heart sound<br>Window length<br>Window shape | Heart sound signals, phonocardiography (PCG) signals, allow for the automatic diagnosis of potential cardiovascular pathology. Such classification task can be tackled using the bidirectional long short-term memory (biLSTM) network, trained on features extracted from labeled PCG signals. Regarding the non-stationarity of PCG signals, it is recommended to extract the features from multiple short-length segments of the signals using a sliding window of certain shape and length. However, some window contains unfavorable spectral side lobes, which distort the features. Accordingly, it is preferable to adapt the window shape and length in terms of classification performance. We propose an experimental evaluation for three window shapes, each with three window lengths. The biLSTM network is trained and tested on statistical features extracted, and the performance is reported in terms of the window shapes and lengths. Results show that the best performance is obtained when the Gaussian window is used for splitting the signals, and the triangular window competes with the Gaussian window for a length of 75 ms. Although the rectangular window is a commonly offered option, it is the worst choice for splitting the signals. Moreover, the classification performance obtained with a 75 ms Gaussian window outperforms that of a baseline method.<br> |

*Corresponding Author:*

Mahmoud Fakhry
Electrical Engineering Department, Faculty of Engineering, Aswan University
81542 Aswan, Egypt
Email: m.fakhry@aswu.edu.eg

## 1. INTRODUCTION

The development of reliable and dynamic noninvasive systems is essential to the successful automatic diagnosis of diseases. Cardiologists are accustomed to using a medical stethoscope to listen to the heart and assess its health. Although this primitive diagnosis strategy is low-cost and non-invasive, its accuracy relays on the hearing experience of the cardiologist, which requires years to acquire [1]. As a result, it sparked the idea of using phonocardiography (PCG) signals to conduct an automated check of heart health [2]–[5].

There are several methods for the automation of heart monitoring and examination, including electrocardiogram (ECG), which records heart electrical activity, and PPG, which estimates the blood flow rate using sensors of light. An assortment of techniques have been developed in the literature for computerizing the investigation of ECG and PPG signals [6]–[8]. Besides ECG and PPG signals, the PCG signals, which records the acoustic sound signals created by the heart in a cardiovascular cycle, can be utilized to proficiently screen and evaluate heart wellbeing. The ECG and PCG signals are more coherent than the PPG signal and they also include more information. In comparison to the ECG signals, the PCG signal has a notable benefit in that it records the acoustic features of heart sound signals. These features make it easier to identify murmurs, which are irregularities in PCG signals [9].

PCG signals are recorded in a quiet room, with the PCG transducer, such as a microphone or stethoscope, firmly placed on the chest with a rubber strap. The PCG signal consists of a pair of heartbeats called S1 and S2, and two silent time durations between them as shown in Figure 1. This refers to the times the heart takes to alternate between contraction and closure. The time duration from the conclusion of the heartbeat S1 to the starting of the heartbeat S2 is called the systolic interval, and the diastolic interval is defined as the time duration from the conclusion of the heartbeat S2 to the starting of the heartbeat S1. The presence of disease is associated with additional audible actions that occur during the silent time durations, e.g., murmur is generated when blood continues to flow during the closure of the heart valves. The heartbeat S1 is a large magnitude signal with low frequency components, and the heartbeat S2 is a small magnitude signal with high frequency components [10]. Many methods have been recently developed for diagnosing heart abnormalities based on the analysis of PCG signals [11], [12]. In general, the detection of heart abnormality based on training distinct features is more effective than using training raw PCG signals.

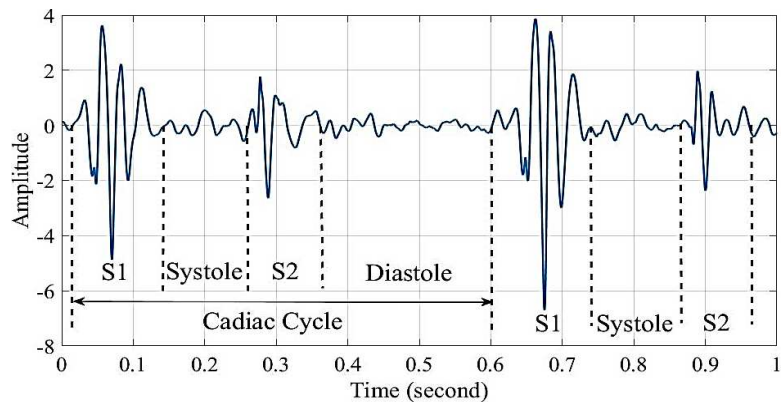

Figure 1. A sound recording of normal heart (PCG signal)

One of the most difficult aspects for extracting features from PCG signals is that their complex and nonstationary nature. They are considered stationary only within short- time intervals, i.e., "quasi-stationary," whereas signal characteristics are nonstationary over longer time intervals. To meet this requirement, features should be extracted in short-time intervals. This means that the PCG signal should be divided into short-length segments of overlapping/non-overlapping signal samples. This is accomplished by employing a sliding temporal window with a specific shape and length [13]. In this sense, the selection of the window shape and its length is a critical issue to ensure a successful diagnosis of heart abnormality.

In this paper, sequences of statistical features extracted from short-length segments of labeled PCG signals are exploited for training and testing the bidirectional long short-term memory (biLSTM) classifier. *Our aim in this work is to experimentally adapt the window shape and length for feature extraction in terms of classification of PCG signals*. For that purpose, the classification results of conducted experiments are reported for three different window shapes and each is investigated with three different lengths. Furthermore, we evaluate four different network architectures of the biLSTM classifier.

The rest of this manuscript is organized. Methods for diagnosing heart diseases are briefly reviewed in section 2. In section 3 explains the proposed method. Elaborated classification experiments and performance are reported in section 4. The conclusion of this work is drawn in section 5.

## 2. LITERATURE REVIEW

Classification features can be extracted from different spectro-temporal representations of PCG signals. These representations can be obtained using the short-time Fourier transform (STFT) [10], the wavelet transform (WT) [14], and the Mel-frequency cepstral coefficients (MFCCs) [15]. Sanei *et al.* [16] conducted a comparative study between these time-frequency representations of PCG signals and the temporal analysis of PCG signals. They have obtained a similar frequency and time resolution for the detection of the acoustic properties of heart sound signals. Because subspaces spanned by heartbeats and murmurs are typically distinct, authors developed a constrained singular spectral analysis method as well as an adapted method for subspace selection to separate murmur from PCG signals in [17]. The optimized singular spectral processing method is found to segregate sound signals even when they are overlapped in time, as evidenced by low correlation values in almost all murmur types. In [18], the support vector machine (SVM) is employed to classify envelopes of PCG signals extracted in the wavelet domain. Hamidi *et al.* [19] perform feature extraction by estimating the parameters of linear predictive coding, and feature classification by SVM with a modified optimizer based on cuckoo search. Curve fitting and fractal dimension are exploited for producing a couple of feature extraction strategies from PCG signals in [20]. In [21], Ölmez and Dokur investigated the classification performance of the k-nearest neighbors (kNNs) with short-time features of wavelet and Hilbert transforms, homomorphic filtering, and power spectrum density of PCG signals. The extracted features are classified employing kNNs with error distances computed using the second norm for multiple cluster numbers.

Artificial neural networks (ANNs) are utilized to classify PCG signals for the first time in [22]. Following that, many scholars focused on diagnosing of heart diseases using various ANNs and features. The principal component analysis and the wavelet decomposition are exploited for extraction of distinct features from PCG signals in [23], [24]. In [25], short-time features of the wavelet analysis and the time-frequency representation of PCG signals were evaluated independently using a multilayer perceptron composed of either a single hidden layer or a couple of hidden layers. The feedforward neural network is used in [26], [27] to classify PCG signals using features computed from the temporal, spectral, and spectro-temporal representations of the signals. The convolutional neural networks (CNNs) have made a rapid progress in the classification of PCG signals [28], [29]. PCG signals are classified using CNNs with features of wavelet coefficients in [30], and of MFCCs in [31]. In [32], heat maps are created from the energy distribution of MFCCs of PCG signals. Authors suggest to combine the heat maps with CNNs to perform such classification tasks.

The recurrent neural networks (RNNs) are combined with MFCCs for the classification of PCG signals. The results are reported for four different RNN models, namely long short-term memory (LSTM), bidirectional LSTM (biLSTM), gated recurrent unit (GRU), and bidirectional GRU (biGRU) [33]. Alam *et al.* [34] describe a combination of two network architectures, namely the CNN and the biLSTM networks. They offer to train visible and temporal properties of murmur in PCG signals through the use of short-time features of spectrogram and MFCCs. The nonlinear autoregressive network with exogenous inputs (NARX) is leveraged for binary the detection of heart abnormality in [35], [36]. This training and testing of this network is performed using three different groups of short-length features, namely, temporal, spectral, and statistical. Comparative analysis of three algorithms of training the biLSTM network for the purpose of PCG signal classification is prepared in [37]. Authors concluded that the network provides the best classification performance when it is trained with the stochastic gradient descent (SGD) with momentum algorithm.

## 3. PROPOSED METHOD

Because PCG signals are complex and nonstationary in nature, their analysis is a difficult issue. They are only deemed stationary within specific short-time intervals. In the proposed method, we offer to split a PCG signal into overlapping short-length segments of samples. This signal dividing step is completed using a sliding symmetric temporal window with certain shape and length. The contribution of each sample of PCG signals in the segment is defined by the window shape, and different segments are obtained by translating the window one sample step. The diagnosing of heart diseases is accomplished by classification of PCG which is achieved using the biLSTM network and 10 statistical features calculated independently inside each segment. However, some window shapes contain unfavorable spectral sidelobes. These sidelobes produce spectral components that interfere with the true spectrum of segments, resulting in distorted features. Using distorted classification features reduces the probability of accurate detection of heart abnormality.

In this proposal, we offer to make a comparison among three different window shapes, each with three different lengths. In this sense, we adapt the shape and length of the window in the context of classification of PCG signals. Sequences of features are employed for training and testing different biLSTM architectures. Figure 2 depicts the flowchart of the proposed method. The sections that follow explain the procedure, beginning with signal pre-processing and progressing to feature extraction and feature classification.

### 3.1. Pre-processing

Most heart sound recordings obtained with recording instruments are corrupted with noise from the surrounding environment. As a result, filtering out the noise becomes critical, but not at the expense of omitting information necessary for diagnosing heart disease. To reject distorting noise from the recordings, PCG signals are severely filtered. The bandwidth of heartbeat S1 is mainly from 10 to 200 Hz, which is made by the vibrations of heart chambers and valves. After the closure of semilunar valves on aortic and pulmonary arteries, the heartbeat S2 is formed with a frequency range from 20 to 250 Hz. The low-frequency components of diagnostic information are preserved while high-frequency noise components are rejected using a low-pass filter with a 250 Hz cut-off frequency. Following Nyquist theorem, the filtered signals are down-sampled before feature extraction to reduce both calculations and computation complexity.

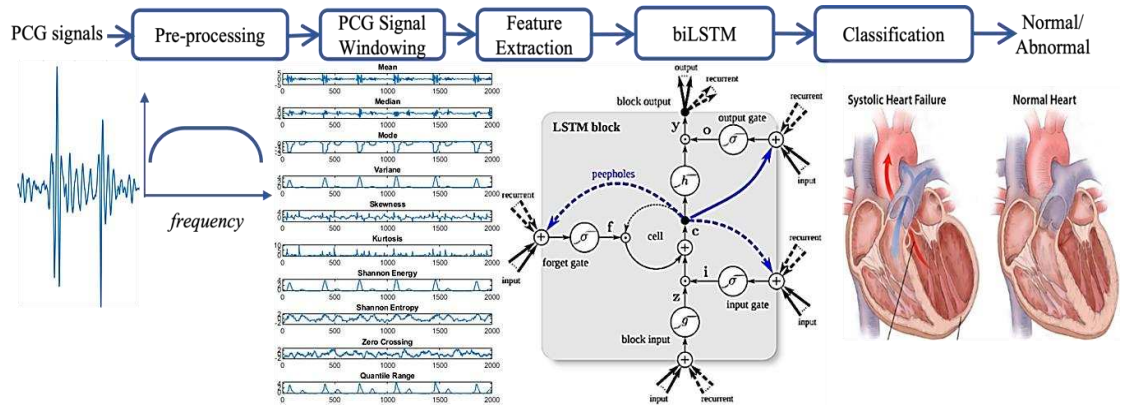

Figure 2. Flowchart of the proposed method

### 3.2. PCG signal windowing

An input signal can be divided into $L+1$ overlapped segments of signal samples. That is, the boundaries of subsequent segments overlap, i.e., the last samples of one segment are present in the following segment. These overlapping segments are obtained by moving a temporal sliding window along the signal. A typical overlap is chosen between one and $L$ samples. In this work, a PCG signal is represented by $N$ overlapping segments of samples with the length of overlap to be $L$. This is achieved for a PCG signal $x[n]$ of length $N$ by employing a sliding symmetric window $w[n]$ of length $L+1$, with $L+1<=N$, such as (1):

$$y_n[l] = w[l]x[n+l], \quad |l| \leq \frac{L}{2}, \tag{1}$$

where $y_n[l]$ is the $n$th signal block of length $L+1$. The shape of the window is selected from widely used windows, including the rectangular window, the triangular window, and the Gaussian window.

#### 3.2.1. Sliding symmetric window

Because the time-domain product of two signals corresponds to the linear convolution of their respective spectra, the spectrum of the signal $x[n]$ is convolved with the spectrum of the sliding window $w[n]$ to obtain the spectrum of $y[n]$. As a result, the spectrum of the window influences the resulting spectrum. The typical spectral shape of a window consists of a main lobe and many side lobes with decreasing intensity as shown in Figure 3. When the signal of interest is sinusoidal, the resulting Fourier transform of the windowed signal will be a superposition of two window functions with their main lobes located at the sinusoidal signal's frequency, effectively "smearing" the delta functions. This unfavorable side effect is usually characterized by [13]: i) the main-lobe width, ii) the intensity of the first (closest) side-lobe peak, and iii) the attenuation of the subsequent side-lobe peaks. Various windows have been offered in the literature to optimize these properties for specific use situations, and some of them are studied here for extracting features in the context of PCG signal classification.

#### 3.2.2. The rectangular window

Because signal truncation corresponds to applying the rectangular window, this is the simplest window. It is also called the Dirichlet window and defined:

$$w[l] = 1, \quad |l| \leq \frac{L}{2}. \tag{2}$$

The spectrum of this window is represented by a *sinc* function, with the peak of the first side-lobe accounting for approximately one-fifth the peak of the main-lobe.

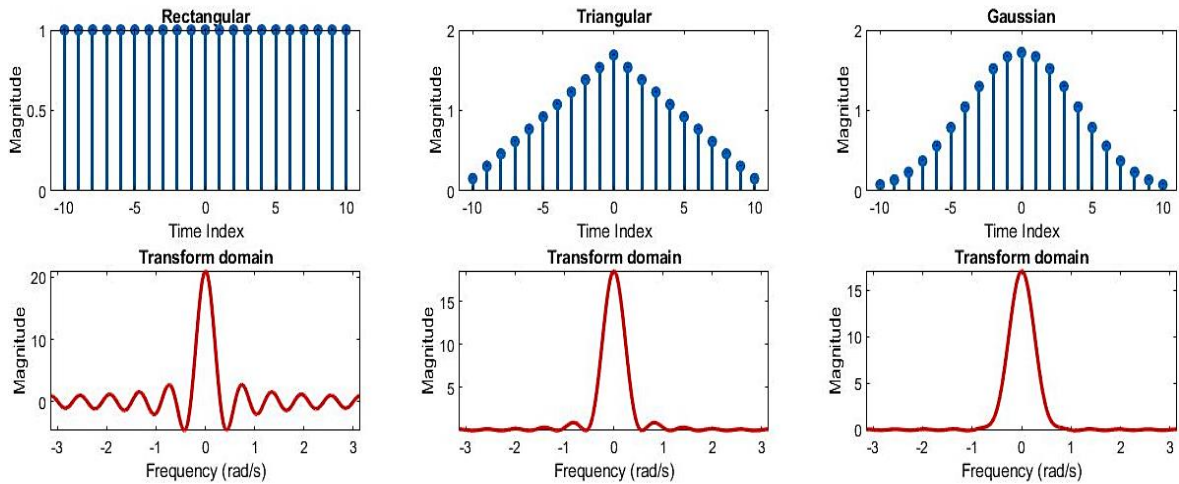

Figure 3. Three different symmetric windows and their corresponding spectral shapes with *L*=20

### 3.2.3. The triangular window

The linear convolution of two rectangular windows of length L/2 yields a (L+1)-length triangular window. Therefore, the spectrum of the triangular window is obviously the Dirichlet kernel squared. This window exhibits a nonnegative Fourier transform due to the above discussed convolution property. Its mathematical formula is determined by (3).

$$w[l] = 1 - \frac{|2l|}{L}, \quad |l| \leq \frac{L}{2}, \tag{3}$$

The half width of the main lobe of the triangular window equals twice that of the rectangular window.

### 3.2.4. The Gaussian window

It is well known that the spectrum of a Gaussian function is also a Gaussian function. Since this function spans an infinity time, it must be trimmed at its ends to be used as a window function. This window is represented by (4).

$$w[l] = e^{-\frac{1}{2}\left(\alpha \frac{l}{L/2}\right)^2}, \quad |l| \leq \frac{L}{2}, \tag{4}$$

Increasing the value of α results in narrowing the width of the window, which reduces the severity of the discontinuity at the edges. However, this will widen the main-lobe and, as a result, diminish the side-lobe levels. The most common values of α are 2.5, 3, and 3.5.

### 3.3. Feature extraction

Various measures for describing the statistical properties of signals are widely used. These measures can be used as classification features extracted from short-length segments of signal samples. Theoretically, the computation of statistical features requires infinite length observations; however, in practical applications, short-length segments can be assumed enough if we adapt the window shape and length [38]. The following section provides explanations of 10 feature sequences representing each PCG signal of the recording database.

### 3.3.1. Arithmetic mean

The arithmetic mean is the average value of the short-length segment of signal samples. Assuming that the samples of a segment are uniformly distributed, the arithmetic mean is computed by (5).

$$\mu_y[n] = \frac{1}{L+1} \sum_{l=-L/2}^{L/2} y_n[l]. \tag{5}$$

For a symmetric probability density function (PDF), the arithmetic mean is the position of the symmetric axis. When the PDF is not symmetric (a skewed signal), then the calculation of the mean is of limited use.

### 3.3.2. Median

The median is the value that spatially divides a segment of samples into half. To find the median, sort the samples of $y_n[l]$ from the smallest value to the largest, and then find the sample with an equal number of samples above and below it. Outlier samples have a minor influence on the median. For a signal with repeated sample values, the median is of limited use, and the mode takes its place as a measure of the average.

### 3.3.3. Mode

The mode is the most frequently occurring value in a segment of samples. In other words, the mode is the common value that is repeated in samples. A multimodal distribution exists when the samples contain multiple values that are linked to the most frequently occurring. If no value repeats, the signal lacks a mode.

### 3.3.4. Variance

The variance and standard deviation both measure the spread of the input segment of signal sample around its arithmetic mean. The difference is given by (6).

$$\sigma_y^2[n] = \frac{1}{L+1}\sum_{l=-L/2}^{L/2}(y_n[l] - \mu_y[n])^2. \tag{6}$$

For $\mu_y[n] = 0$, the variance equals the power of the observed segment of signal samples. The standard deviation $\sigma_y[n]$ can be computed directly from the variance, $\sigma_y[n] = \sqrt{\sigma_y^2[n]}$.

### 3.3.5. Skewness

The skewness is defined as the division of the third central moment of a segment of signal sample by the cube of its standard deviation. It is given by (7).

$$v_y[n] = \frac{1}{L+1}\frac{1}{\sigma_y^3[n]}\sum_{l=-L/2}^{L/2}(y_n[l] - \mu_y[n])^3. \tag{7}$$

Skewness is a measure of PDF asymmetry. It is 0 for symmetric distributions, negative for distributions with their mass centered on the right, and positive for distributions with their mass centered on the left.

### 3.3.6. Kurtosis

Kurtosis is defined as the division of the fourth central moment of a segment of signal sample by the fourth power of the standard deviation. It is mathematically computed as (8):

$$\omega_y[n] = \frac{1}{L+1}\frac{1}{\sigma_y^4[n]}\sum_{l=-L/2}^{L/2}(y_n[l] - \mu_y[n])^4 - 3. \tag{8}$$

Kurtosis quantifies the non-Gaussianity of the PDF. It is an indicator of the flatness of the PDF compared to the Gaussian distribution. It is zero for a Gaussian distribution, is negative for a flatter distribution with a wider peak, and positive for distributions with a sharper peak.

### 3.3.7. Shannon energy

Shannon energy finds the average spectral of a segment of signal samples. It discounts the high value components into the low value components. It is obtained for a segment of samples.

$$Es_y[n] = \sum_{l=-L/2}^{L/2}|y_n[l]|^2 \log|y_n[l]|^2. \tag{9}$$

In the presence of noise and outliers, Shannon energy approaches signal ranges, resulting in fewer errors. The advantage of using Shannon energy is its ability to emphasis the medium over traditional energy.

### 3.3.8. Shannon entropy

Shannon entropy quantifies the uncertainty of a random variable. It is calculated using the PDF $p_y(y_n[l])$ of a segment of samples.

$$hs_y[n] = \sum_{l=-L/2}^{L/2} p_y(y_n[l]) \log p_y(y_n[l]). \tag{10}$$

Entropy measures how effectively one can predict the behavior of respective parts from the others. In general, more entropy denotes more complicated signals and, as a result, less predictability.

### 3.3.9. Zero-crossing rate

The number of sign changes in subsequent samples is referred to as the zero-crossing rate. It is a low-level feature that has been utilized in audio analysis for decades due to its straightforward calculation.

$$z_y[n] = \frac{1}{2}\frac{1}{L+1}\sum_{l=-L/2}^{L/2}|\text{sign}(y_n[l]) - \text{sign}(y_n[l-1])|. \tag{11}$$

The output is in the range of $0 \leq z_y[n] \leq 1$. The more often the signal changes its sign, the more high-frequency content is assumed to be in the signal.

### 3.3.10. Quantile ranges

Quantiles can be used to split the PDF into subsets of equal size. When the PDF is divided into two quantiles, it is divided into two halves, each of which contains half of the total number of observations. The median will be defined as the point where the border between those two quantiles crosses.

$$Q_y(0.5) = y|_{\int_{inf}^{y} p_y(x)dx = 0.5} \tag{12}$$

In the case of symmetric distributions, this equals the arithmetic mean. Quantile ranges are useful for simplifying the explanation of a distribution's shape. Here, we adopt the range spanned by 50 % of the samples by discarding the upper and lower 25 %. The quantile range is then computed.

$$\Delta Q_y(0.5) = Q_y(0.75) - Q_y(0.25). \tag{13}$$

### 3.4. Normalization of feature sequences

Normalization rescales the features so that they have a zero mean and unit standard deviation. For a feature vector $a[n]$ with mean $\mu_a$ and standard deviation $\sigma_a$, the normalization is accomplished.

$$a[n] = \frac{a[n]-\mu_a}{\sigma_a}. \tag{14}$$

Normalization of features to be centered around zero with a standard deviation of one is important for comparing measurements with different units, and it is also required by many learning algorithms. This normalization is applied to all of feature sequences obtained as shown in the previous section. Figure 4 depicts an example of 10 normalized sequences of extracted features from two PCG signals. The normalized sequences are then used to train and test the biLSTM network.

### 3.5. Feature classification

Recurrent neural networks (RNNs) [33] are composed of hidden layers and feedback connections. In RNN models there are at least recurrent layer with numerous hidden neurons/nodes between input and output layers. The recurrent long short-term memory (LSTM) network contains hidden layers with self-recurrent weights, allowing memory to retain previous information. This network is well-known for modeling the trend in a sequence of features [39]. The bidirectional LSTM (biLSTM) network is a modified LSTM network with two hidden layers. These hidden layers learn sequences by alternating between backward and forward layers [40]. The alternating strategy improves the accuracy of LSTM networks by enabling past data to offer context for subsequent samples in a training sequence.

### 3.5.1. Training the biLSTM model

Backpropagation is applied to calculate the training parameters of the biLSTM network, which is then followed by an optimization algorithm [41]. The simplest way is to take each training example, run it through the network to get a prediction output, subtract it from the actual output we want to get, and square it. The network performs very well if the loss function is small, and it should be as small as possible. Backpropagation computes the gradients used by the optimization algorithm to minimize loss functions. To update the network parameters, we use the stochastic gradient descent (SGD) with momentum. Unlike the SGD, the momentum considers the gradient obtained in the last steps to find the optimum search path. This allows the network to approach the minimum value of the loss function more quickly. Although the momentum speeds up convergence, it should be used in conjunction with simulated annealing to avoid overshooting global minima. In the real world, the momentum is set at values that start at 0.5 and progressively increase to 0.9 over the course of the training epochs.

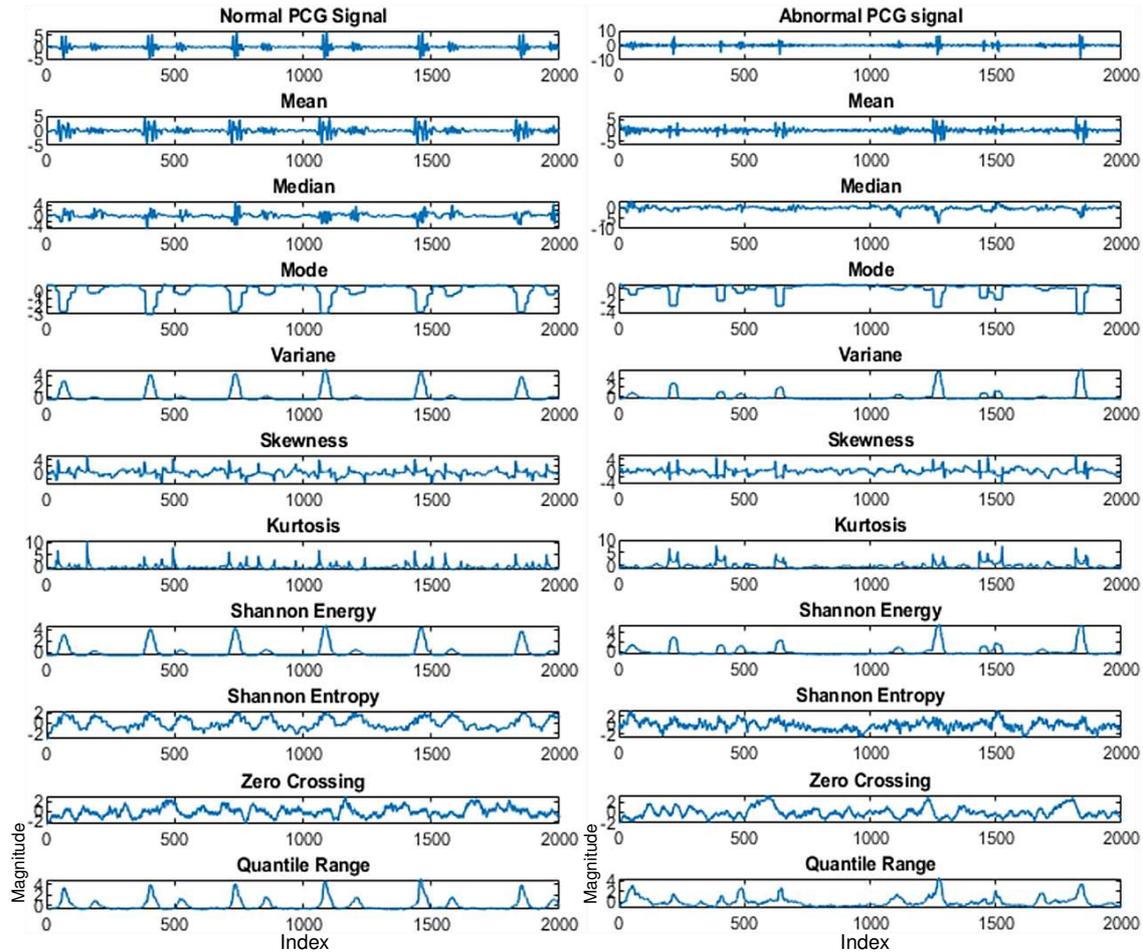

Figure 4. Normalized sequences of features for two PCG signals obtained with a 75 ms Gaussian window

## 4. EXPERIMENTS
### 4.1. Database
The heart sound recordings considered here are taken from the database of the PhysioNet 2016 challenge, which is publicly available on the Web [42]. An electronic stethoscope was used to record PCG signals at four different points on the chest. The database includes recordings that have been classified as healthy or pathological. We chose a well-balanced database that includes sound recordings of 150 healthy hearts and other 150 pathological hearts. These recordings range in length between 5 and 120 seconds, with a sampling frequency of 2,000 Hz. Following Nyquist theorem, all recordings are downsampled at 500 Hz.

### 4.2. Training and testing
For extraction of short-time features, we truncate the downsampled PCG signals, so that their lengths become 10 seconds (5,000 samples). The truncated signals are divided into short-length segments of samples using a window of length $L+1$. 70% of the extracted features are used to train the model, while 30% are used to test it. The training and testing sets are picked from the extracted features of the recording database at random. This procedure of training and testing is done for thirty times separately. Averaging the outcomes of the thirty experiments yields the classification performance. The classification results are provided in terms of the shape and length of the sliding symmetric window and as a function of four different numbers of the hidden neurons of the biLSTM classifier. We set the value of the model variables as listed in Table 1. These variables include the learning rate and the value of momentum in SGDM.

### 4.3. Classification performance
The number of successfully classified signal examples vs the number of wrongly classified signal examples is a measure of classification performance. False negatives (FN) occur when the total number of signals from pathological patients are reported healthy, while false positives (FP) occur when the total number of signals from healthy patients are proclaimed pathological. True positives (TP) refer to pathological heart sounds that have been accurately identified, whereas true negatives (TN) are healthy heart sounds that have been appropriately categorized. The sensitivity (Sens.), specificity (Spec.), and accuracy (Accu.) are commonly computed using the above-defined variables [43].

$$Sens. = \frac{TP}{TP + FN} 100\%,$$

$$Spec. = \frac{TN}{TN + FP} 100\%$$

$$Accu. = \frac{TP + TN}{TP + FN + TN + FP} 100\%.$$

Table 1. Implementation variables of the biLSTM classifier with SGDM

| Variable | Value |
|---|---|
| # biLSTM layers | 2 |
| # Neurons in layer | 5, 30, 50 or 100 |
| # Classes | 2 |
| Initial learning rate | 0.01 |
| Epochs | 500 |
| The momentum | 0.90 |

### 4.4. Results

Figure 5 depicts the classification records for the selected test dataset averaged over the number of hidden neurons in the biLSTM model. The best performance is obtained by employing the Gaussian window with a length of 30 samples (75 ms) to split PCG signals into short-length segments for feature extraction, followed by the triangular window. Furthermore, the specificity and the accuracy accomplished by the classifier are better when short-length segments are obtained using the triangular window of length 50 samples (125 ms) than those by the Gaussian window, and the sensitivity values are comparable. The results obtained with the Gaussian window surpass those with the triangle and rectangular windows in terms of sensitivity and accuracy for windows of length of 15 samples (37.5 ms), while the results with the rectangular window are better than those with the other windows in terms of specificity, followed by those with the Gaussian window.

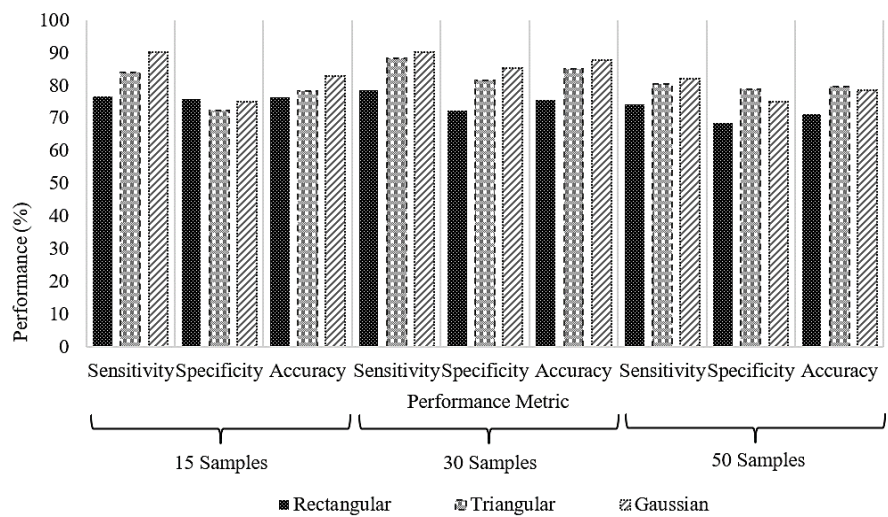

Figure 5. Average performance for four numbers of neurons of the classification model

Figure 6. depicts three sub-figures of the accuracy obtained by the biLSTM network. Each sub-figure shows the accuracy when each window of the three windows is used to divide PCG signals into short-length segments. The accuracy follows different trends for different numbers of hidden neurons.

Overall window shapes, the optimal accuracy achieved by the biLSTM classifier when the number of its hidden neurons is set to 30. For a large or small number of hidden neurons, accuracy degradation is recorded. When using the biLSTM network to classify features extracted from short-length segments of PCG signals with a sliding symmetric window, the Gaussian window or the triangular window is the best choice. However, the rectangular window outperforms the other two windows for a length of 15 samples (37.5 ms).

Table 2 reports detailed classification performance for the selected test PCG signals. According to the results, using any window length and shape provides a good identification of heart abnormality for all numbers of hidden neurons in the biLSTM classifier, as evidenced by the sensitivity values compared to the specificity. The window shape and length can be adapted to get the best performance. The best achieved accuracy is 89.10.

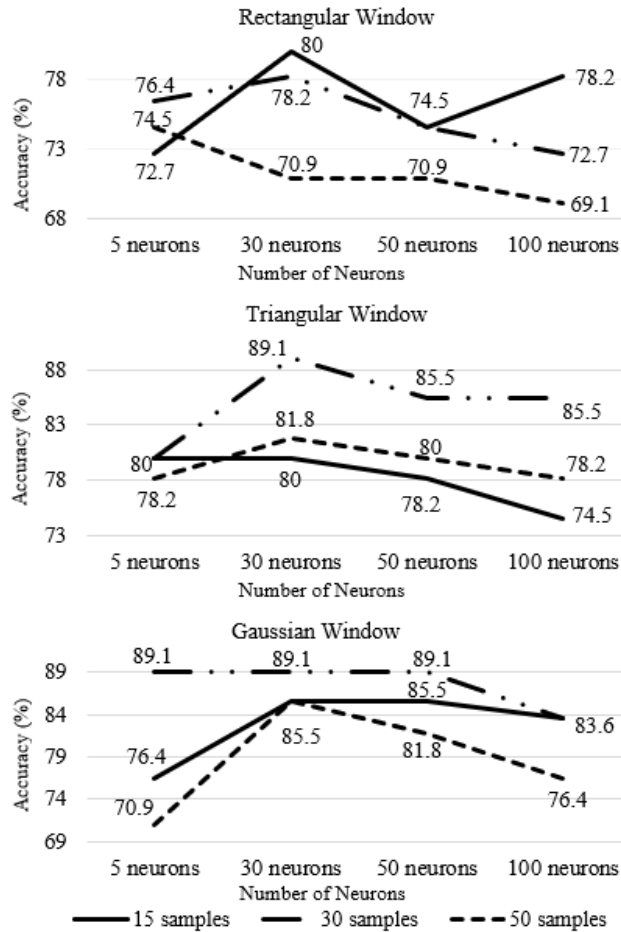

Figure 6. Classification performance of the test dataset in terms of accuracy for each window shape as a function of a different number of hidden neurons and different window lengths

Table 2. Detailed classification results for the test data

| Window length | | 15 samples | | | 30 samples | | | 50 samples | | |
|---|---|---|---|---|---|---|---|---|---|---|
| Window shape | Neurons | Sens. | Spec. | Accu. | Sens. | Spec. | Accu. | Sens. | Spec. | Accu. |
| Rectangular | 5 | 71.40 | 74.10 | 72.70 | 78.60 | 74.10 | 76.40 | 78.60 | 70.40 | 74.50 |
| | 30 | 82.10 | 77.80 | 80.00 | 85.70 | 70.40 | 78.20 | 71.40 | 70.40 | 70.90 |
| | 50 | 75.00 | 74.10 | 74.50 | 75.00 | 74.10 | 74.50 | 71.40 | 70.40 | 70.90 |
| | 100 | 78.60 | 77.80 | 78.20 | 75.00 | 70.40 | 72.70 | 75.00 | 63.00 | 69.10 |
| Triangular | 5 | 82.10 | 77.80 | 80.00 | 85.70 | 74.10 | 80.00 | 75.00 | 81.50 | 78.20 |
| | 30 | 89.30 | 70.40 | 80.00 | **92.90** | **85.20** | **89.10** | 85.70 | 77.80 | 81.80 |
| | 50 | 85.70 | 70.40 | 78.20 | 85.70 | 85.20 | 85.50 | 82.10 | 77.80 | 80.00 |
| | 100 | 78.60 | 70.40 | 74.50 | 89.30 | 81.50 | 85.50 | 78.60 | 77.80 | 78.20 |
| Gaussian | 5 | 85.70 | 66.70 | 76.40 | **89.30** | **88.90** | **89.10** | 75.00 | 66.70 | 70.90 |
| | 30 | 92.90 | 77.80 | 85.50 | **92.90** | **85.20** | **89.10** | 89.30 | 81.50 | 85.50 |
| | 50 | 92.90 | 77.80 | 85.50 | **92.90** | **85.20** | **89.10** | 85.70 | 77.80 | 81.80 |
| | 100 | 89.30 | 77.80 | 83.60 | 85.70 | 81.50 | 83.60 | 78.60 | 74.10 | 76.40 |

*Comparison of window shapes and lengths in short-time feature extraction for … (Mahmoud Fakhry)*

### 4.5. Discussion

Adaptation of the shape and length of sliding window to split PCG signals into short-length segments for feature extraction is an intriguing need to improve the diagnosis accuracy of heart disease. The nonstationary nature of these signals and the presence of side-lobes in the spectrum of windows encourage us to search for optimal window length and shape to avoid destructive effects on the performance, as is theoretically demonstrated in the literature and experimentally confirmed in this work. Using a sliding window with adapted length and negligible side-lobes, such as the Gaussian window, provides improved performance.

To demonstrate the importance of this work, we compare the results obtained by the proposed method to those obtained by a baseline method of [21]. In this baseline method, authors offer to extract four sets of features obtained from the wavelet and Hilbert transforms, the homomorphic filtering, and the power spectral density. This results in eleven extracted features, which were classified using kNNs and yielded classification accuracy ranging from 74.07 to 81.40. These values of accuracy outperform many other classification methods, as reported in [21]. The proposed method achieves a classification accuracy of 89.10 with the Gaussian or triangular windows of length 30 samples (37.5 ms) for extracting features and setting the number of hidden neurons to 30 for classifying features. In this regard, the proposed method outperforms the baseline method and achieves accuracy gain ranging between 15.03 and 7.70.

## 5. CONCLUSION

In this paper, we proposed to optimize the shapes and the lengths of sliding symmetric temporal windows for splitting PCG signals into short-length segments, for extraction of sequences of statistical features. We used these sequences to train and test the biLSTM classifier for the purpose of diagnosing heart abnormalities. We tested three different window shapes, each with three different window lengths for the extraction of features from short-length segments and four different numbers of hidden neurons for feature classification. According to our findings, the classification performance obtained using the triangular window is comparable to that obtained using the Gaussian window. However, the performance obtained using the rectangular window cannot compete with that obtained using the other two windows. This results in the high-level sidelobe present in the spectrum of the rectangular window compared to the spectra of the other windows. These sidelobes interfere with the spectrum of signal segment. The best classification performance is achieved when we used the Gaussian or triangular window of length 75 ms to extract features from short-length segments of PCG signals, and we set the number of neurons of the biLSTM classifier to 30. With this set of variables, the proposed methodology outperforms a baseline classification method.

## BIOGRAPHIES OF AUTHORS

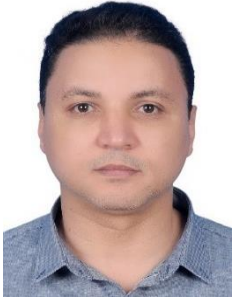 **Mahmoud Fakhry** received the Ph.D. degree of ICT from the University of Trento, Italy, in 2016. He worked for the Center for Information and Communication Technology, Fondazione Bruno Kessler, Trento, Italy from October 2012 to November 2016. From December 2016 to February 2018, he was a researcher fellow at the Audio Analysis Lab, Aalborg University, Denmark. From March 2018, he is an assistant professor at the Electrical Engineering Department, Aswan University, Egypt. His research interests include audio signal analysis and machine learning. He can be contacted by email: m.fakhry@aswu.edu.eg.

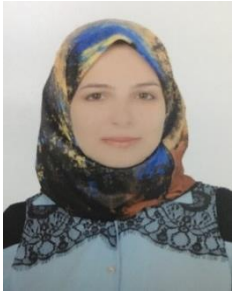 **Abeer FathAllah Brery** received the master's degree in computer science and engineering from Universitat Rovira i Virgili, Tarragona, Spain, in September 2016. From October 2016, she is a research and teaching assistant at Electrical Engineering Department, Aswan University, Aswan, Egypt. Her research interests include software engineering and machine learning. She can be contacted at email: abeer_brery@aswu.edu.eg.